\documentclass[aps,prl,reprint,showpacs,floatfix,groupedaddress]{revtex4-1}
\usepackage{multirow}
\usepackage[T1]{fontenc}
\usepackage[cp1250]{inputenc}
\usepackage{psfrag}
\usepackage{graphicx}
\usepackage[english]{babel}
\newcommand{\bq}{\begin{equation}}
\newcommand{\eq}{\end{equation}}
\newcommand{\ba}{\begin{eqnarray}}
\newcommand{\ea}{\end{eqnarray}}
\begin{document}

\title{Temporal self-similar synchronization patterns and scaling in repulsively coupled oscillators}


\author{Darka Labavi\'c}
\author{Hildegard Meyer-Ortmanns}
\email{h.ortmanns@jacobs-university.de}
\affiliation{Physics and Earth Sciences, Jacobs University, P.O. Box 750561, 28725 Bremen, Germany}




\begin{abstract}
We study synchronization patterns in repulsively coupled Kuramoto oscillators and focus on the impact of disorder in the natural frequencies. Among other choices we select the grid size and topology in a way that we observe a dynamically induced dimensional reduction with a continuum of attractors as long as the natural frequencies are uniformly chosen. When we introduce disorder in these frequencies, we find limit cycles with periods that are orders of magnitude longer than the natural frequencies of individual oscillators. Moreover we identify sequences of temporary patterns of phase-locked motion, which are self-similar in time and whose periods scale with a power of the inverse width about a uniform frequency distribution. This behavior provides challenges for future research.
\end{abstract}







\maketitle

\section{Introduction}
As model for classical oscillators we consider the Kuramoto model, a paradigm for synchronization phenomena with applications to natural and artificial systems \cite{acebron}. Although in natural systems like neuronal or genetic networks one often finds antagonistic couplings (excitatory and inhibitory, attractive and repulsive), numerous modifications of the Kuramoto model were studied for only attractive couplings, and if repulsive ones were included, they were assigned randomly to the links of the grid \cite{zanette,toral}. The interplay of antagonistic couplings usually enriches the attractor space. In particular, if the combination of repulsive and attractive couplings with a certain grid topology leads to unsatisfied bonds and therefore to frustrated interaction loops \cite{daido} in the system, one expects multistable behavior in analogy to spin systems, where frustrated loops are responsible for a high degeneracy of the ground state and a rough energy landscape.

Under the additional action of noise further analogies to spin systems can be observed. If stochastic fluctuations are induced by noise and ``order" refers to coalescing phases in synchronization patterns, we observed an order-by-disorder phenomenon in repulsively coupled Kuramoto systems and classical rotators \cite{mischa}. Order-by-disorder phenomena are also known form spin systems, when stochastic fluctuations are induced by a finite temperature or the quantum nature of the system, and order refers to the magnetization degree of the spins (see, for example, \cite{orderbydisorder}). In \cite{mischa} we found already first hints for a rather rich attractor space with a hierarchy in the potential barriers. Differently from coherence resonance \cite{arkady} an increasing order was not only found for an intermediate noise strength, but noise intervals were identified that led to alternating order and disorder in the oscillator phases. When noise is turned on, the attractors become metastable states, from which the oscillator phases are kicked from one such state to the next under the action of noise, with a variety of escape times. These observations led us to the expectation to observe ``physical aging" in these oscillatory systems. The notion of physical aging is much more restricted
than biological aging. It refers to the breaking of time translation invariance in the measurement of autocorrelation and response functions, in which the system depends on its ``age" after a quench, see, for example \cite{pleimling}. In our oscillatory systems the quench refers to a choice of initial conditions from the monostable phase, after which the coupling is chosen from the multistable regime. The expectation was confirmed in \cite{florin} and raised the question about the role of noise in physical aging. In \cite{chaos} we therefore replaced the noise by a distribution in the natural frequencies. Again we observed physical aging, but the mechanism of how the system explores the rich attractor space via deterministic trajectories is quite different from a migration of phases under the action of noise. As it turned out,  what is responsible for aging is the variety of long transient times until the system finds its stationary state. The stationary states can be limit cycles with periods of the order of the inverse natural frequencies, but also orders of magnitude longer \cite{chaos}. Moreover, we found first hints for self-similar sequences of synchronization patterns within the long-period orbits, when zooming in in time and accordingly scaling the width of deviations from the uniform frequency distribution.

In this paper we further explore the self-similarity of temporary sequences and their scaling properties and extend the analysis from the formerly chosen hexagonal grids to ring topologies.
As we shall see, the scaling range and power law depend on the grid topology and its size. 

The paper is organized as follows. In section 2 we specify the model and summarize some background information about the attractor space for uniform and disordered natural frequencies. Section 3 provides the results on self-similar patterns and scaling properties for hexagonal grids and ring topologies, followed by a stability analysis of the long-period orbits. In section 4 we summarize our conclusions with an outlook to further challenges.

\section{The model and its rich attractor space}\label{sec2}
\subsection{The model}
We consider systems of $N$ Kuramoto oscillators~\cite{kuramoto},
whose phases $\phi_i$ are governed by the equations:
\begin{equation}
\label{eq1}
\frac{d\phi_i}{dt} =
\omega_i +  \frac{\kappa}{\mathcal{N}_i}\sum_j A_{ij}\sin(\phi_j-\phi_i).
\end{equation}
Here $\omega_i$ denote the natural frequencies,
$\kappa$ parameterizes the coupling strength, chosen to be negative throughout the paper, that is in the regime of multistability. Further, $\mathcal{N}_i$ denotes the number of neighbors to which the $i$-th unit is connected, and $A_{ij}$
is the adjacency matrix with $A_{ii}=0$, $A_{ij}=1$ if
$i\neq j$ and units $i$ and $j$ are connected, otherwise $A_{ij}=0$.
We choose $A_{ij}$ to represent either a hexagonal lattice of the size
$L\times L$ with periodic
boundary conditions, so that $\mathcal{N}_i=6$ for all sites $i$,  or a ring $(R,N)$ of N nodes, undirected and with degree R, so that each node is coupled to $R/2$ neighbors to the right and to the left.
The peculiar choice of a hexagonal lattice together with exclusively repulsive couplings guarantees a rich attractor space in a controlled way, as each elementary triangle contains at least one unsatisfied bond, so that frustrated triangles induce a proliferation of attractors. Multistability for the ring topology is also expected, because we consider choices of N and R with a number of frustrated loops. Loops with an odd number of undirected repulsive couplings contain at least one unsatisfied bond.

\begin{figure}[!t]
\centering{
\includegraphics[width=.9\columnwidth]{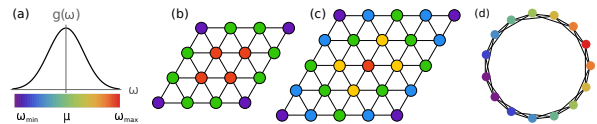}
}
\caption{Regular assignment of natural frequencies for two sizes $l_x\times l_y = 4\times4,  5\times5$ of a hexagonal lattice and a ring (4,15) of 15 nodes with degree 4, connected to two next neighbors and two next-to-next neighbors on both sides of a node. The colors represent the values of $\omega_i=\omega+\delta \omega_i$, orange for the largest deviation $\delta\omega_i$, and purple for the smallest.}\label{fig1}
\end{figure}

For the frequency distributions we choose a regular implementation to easily control the ``distance" to the uniform frequency distribution.
We generate frequencies according to a deterministic procedure, the formulas can be found in \cite{chaos}. The choice is most easily illustrated in Fig.~\ref{fig1}.
The frequencies are uniformly spaced with respect to a Gaussian
\begin{equation} g(\omega)=\frac{1}{\sqrt{2\pi\sigma^2}}e^{-(\omega-\mu)^2/(2\sigma^2)}
\end{equation}
with mean $\mu=1$ and width $\sigma$. This generates  ``spatial" Gaussians in natural frequencies with the largest frequencies in the center of the displayed hexagonal grid (e.g., four sites in the middle of the $4\times 4$ grid of Fig.~\ref{fig1}), or at some site on the ring with isotropic gradients from the maximal frequency deviations $\delta \omega_i$ from $\omega=\mu=1$ towards smaller ones. This way the deterministic system can be smoothly tuned towards the uniform frequency $\mu=\omega$ for all sites by varying $\sigma$. We shall analyze the dependence on the initial conditions for $\phi_i$, keeping the distribution of $\omega_j$ fixed in a given run.

\subsection{A dynamically generated Watanabe-Strogatz phenomenon}
By Watanabe-Strogatz (WS)-phenomenon we refer to the results of Watanabe and Strogatz \cite{watanabe} on a possible dramatic dimensional reduction that happens in a system of N identical one-dimensional elements
\begin{equation}
\frac{d\phi_i}{dt}\;=\;f(\phi_i)+\kappa\sum_j A_{ij} g(\phi_,\phi_j)
\end{equation}
with local dynamics described by $f(\phi_i)$, adjacency matrix $A_{ij}$, describing global coupling topology and $g(\phi_i,\phi_j)$ only including first Fourier harmonics. The prediction then  is that there are $N-3$ conserved quantities. In particular, under the former conditions any system of N sinusoidally coupled phase oscillators can be reduced to a three-dimensional system with $N-3$ constants of motion. It should be noticed that the coupling between the N identical elements should be global. 

First in \cite{mischa} we observed a dynamically generated WS-phenomenon: 
Starting from a certain subset of initial conditions, the sixteen oscillators on the hexagonal grid arranged in four clusters, which are globally coupled when each cluster is represented by a collective variable. The reduced system of four variables satisfies the conditions for the WS-phenomenon to occur with N-3=1 conserved integral of motion.The continuum of solutions differs by their frequencies, as it was first observed and identified by M. Zaks.  The global coupling is dynamically generated, as we started from local hexagonal couplings when the four clusters formed.

In \cite{tomov} general necessary conditions were analyzed, which allow the WS-phenomenon in networks with non-trivial coupling topology. In particular it is expected for special ring topologies, of which we consider two cases, undirected rings $(4,15)$ and $(6,12)$. The $(4,15)$ system arranges in particular into five globally coupled clusters of three oscillators each, while the $(6,12)$ system into four clusters of three oscillators each. A detailed stability analysis for these states can be found in \cite{tomov}. Even if states differ just by permutations of phases, their embedding in the higher-dimensional manifold, spanned by the original variables, may be different and change the stability properties of these permutations. It should be mentioned that the continua of WS-attractors in these systems are in general not the only stable solutions of phase-locked motion, but if they are there, the multistability is extreme \cite{feudel}, so that effects related to multistability should be rather pronounced. This is the reason why we perturb the natural frequencies about a uniform distribution primarily for systems, which allow the dynamical WS-phenomenon. As it turned out, however, the emergence of long-period orbits, their self-similarity and scaling are not restricted to these particularly rich attractor spaces, as we shall see in section 3.

\subsection{The impact of disorder in natural frequencies on the attractor space}
\begin{figure}[!t]
\centering{
\includegraphics[width=1.0\columnwidth]{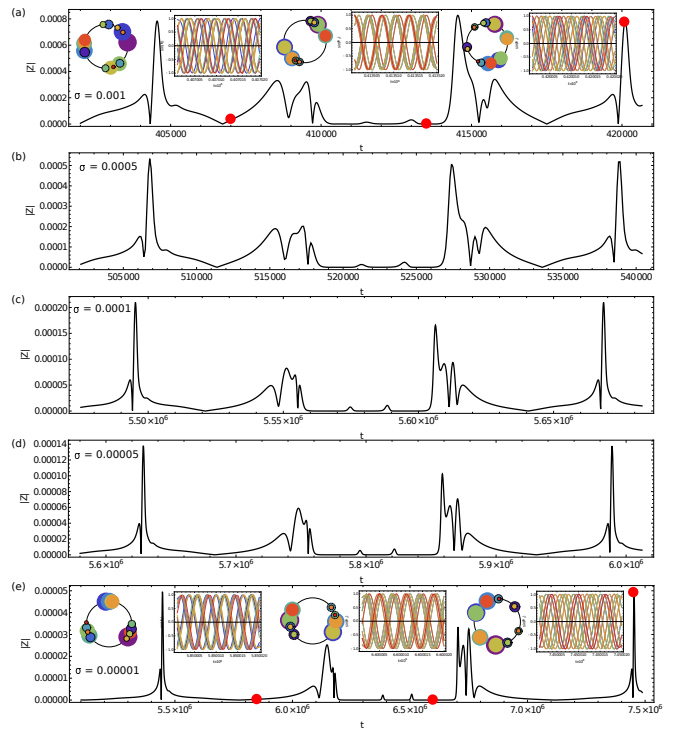}
}
\caption{Self-similar orbits of an order parameter for a 4x4 hexagonal lattice for five different values of $\sigma$ a) 0.001, b) 0.0005, c) 0.0001, d) 0.00005, and e) 0.00001, with periods T = 15555, 32031, 173086, 360200, and 199060 t.u., respectively. Insets in figures a) and e) show phase evolutions for 20 time units (t.u.), starting at time instants, indicated by red dots on the order parameter orbits $\vert Z\vert$, and corresponding phase representations on the unit circle.  It should be noticed that the time evolutions of  $\vert Z\vert$ along with the phase evolutions are self-similar, when $\sigma$ is varied, while the long periods considerably change between a) and e).}\label{fig2}
\end{figure}

If we now turn on a distribution in the natural frequencies $\omega_i=\omega+\delta \omega_i$ and start from different initial conditions for a given fixed distribution, we observe different type of trajectories. Apart from limit cycles, whose periods are of the order of the inverse natural frequency, we find typically a number of long-period orbits, as displayed in Fig.~\ref{fig2} for a $4\times 4$ hexagonal grid at $\kappa=-2$. We present the long-period orbit in three representations.
Most convenient for identifying long-period orbits is a plot of the Kuramoto order parameter, defined as
\begin{equation}
Z(t):= \vert Z\vert e^{i\theta}\;=\;\frac{1}{N}\sum_{j=1}^N e^{i\phi_j},
\end{equation}
where a positive value of $\vert Z\vert$ in the limit of $N\rightarrow\infty$ implies the emergence of phase synchronization and $\theta$ is the phase of the global order parameter \cite{kuramoto}.
It is first $\vert Z\vert$, for which we observe the striking long periods, by orders of magnitude longer than individual periods of oscillators, in the displayed case the periods are minimally 15555 t.u. for the maximal width and maximally 1990600 for the minimal width. Although identical values of $\vert Z\vert$ are compatible with, but not conclusive for the same patterns of phase-locked motion,  it is a useful first tool to record and register these long periods. Long periods of the trajectories may otherwise be easily overlooked when following the individual time evolution of larger sets of oscillators. The actual phase trajectories of the sixteen phases in Fig. \ref{fig2} do have the same period as $\vert Z\vert$ and show a striking self-similarity that will be the topic of the next section. For the self-similar time evolutions of the synchronization patterns it should be noticed that the smaller the variance $\sigma$, the larger the period of the system.

At a few selected instants of time we also show the actual phase trajectories over a time interval of 20 time units. Obviously the synchronization patterns of phase-locked motion are temporary and can considerably differ between different instants of time. So it may come as a surprise that nevertheless, after a long period, the patterns exactly repeat. Obviously the phases only approximately arrange into well defined clusters, and the change of the state is very smooth and can take several individual periods even for the shortest long periods.

The period of an individual oscillator is approximately $2 \pi /\omega_i$, where $\omega_i = \omega + \delta \omega_i$, with $\omega=1$ being the mean of the normal distribution, and $\delta\omega_i<<\omega$ depends on the variance $\sigma$ of the normal distribution. In spite of the small change in $\omega_i$, the long period of the system depends strongly on  $\sigma$. For a larger long-period, the system stays longer in a certain state, compare the approximate three-cluster state in figure a), where it lasts $< 2000$ t.u., while in e) $\sim 350000$ t.u.. More about the scaling relation between the long-period T with $\sigma$ will follow in figure~\ref{fig3} below.

The third type of representation of phases is via discs on the unit circle. Each of the sixteen phases is represented as a colored disc with a specific diameter. The only reason for the varying diameter is to make the phases distinguishable on the circle even if all phases coalesce. Moreover it can be easily read off what type of approximate cluster state the system is in, for example that a temporary repetitive pattern is an approximate three-cluster state (most left in a) and e)), or a two-cluster state in the middle of a) and e).

\section{Results}

\subsection{Hexagonal grids in two dimensions}
We consider two hexagonal grids, $4\times 4$ and $5\times 5$, the $4\times 4$-grid allows a WS-dimensional reduction for $\sigma=0$ or $\omega_i=\mu$ in contrast to the $5\times 5$-grid, where a globally coupled arrangement of clusters is not compatible with the grid size. In both systems we select long-period orbits with repeating patterns of phase-locked motion and first check whether we find similar sequences of patterns once $\sigma$ is varied. The similarity is recorded by the time evolution of $\vert Z\vert$ as in Fig.~\ref{fig2}. When it is visualized via the individual phases,
the characterization in terms of temporary clusters sometimes is marginal, since the phases within one cluster do not agree within the numerical accuracy, but within some tolerance interval such that they  differ less from other phases inside than outside the same cluster. The question may arise as to whether we can find self-similar orbits to any identified long-period orbit. Since a complete overview of the attractor space is not possible, we can only state that we have not found any counterexamples so far.

The long periods are determined by peak detection. We compare a few maximum peaks: If they have the same value and repeat after the same time interval, we record this time interval as the period $T$. In general we include data for the period only for those values, for which we are sure that the system evolved indeed to a similar orbit attractor, i.e. we can determine the period within the accuracy of one t.u., which is 100 integration steps. 

\begin{figure}[!t]
\centering{
\includegraphics[width=1.0\columnwidth]{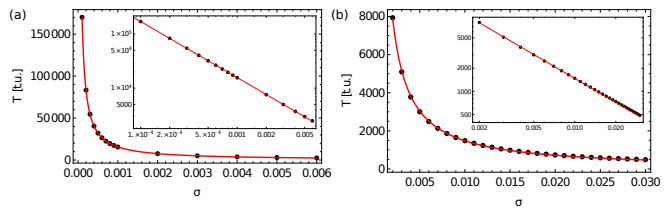}
}
\caption{Periods of the order parameter $\vert Z\vert$ along selfsimilar orbits as a function of the width $\sigma$ of the distribution of natural frequencies for a $4\times4$ (a), and $5\times5$ (b) hexagonal grid. Black dots represent data, red curves power law fits $T(\sigma) = a \sigma^b$ with $a = 12.2\pm0.1$ and $b=-1.0366\pm0.0009$ (a) and $a = 12.4\pm0.2$ and $b=-1.038\pm0.003$ (b). The insets show the same data in a log-log scale. $\sigma$ ranges from 0.0001 to 0.006 (a), and
from 0.002 to 0.03 (b).}\label{fig3}
\end{figure}

A precise measurement of the long-period is important, if we want to check the scaling behavior of the period with the distance from the uniform frequency distribution, parameterized by $\sigma$, as shown in Fig.~\ref{fig3} for both grid sizes. Black dots represent data, the red curves are power law fits $T(\sigma) = a \sigma^b$, where $a = 12.2\pm0.1$ and $b=-1.0366\pm0.0009$ ($4\times4$-grid), and $a = 12.4\pm0.2$ and $b=-1.038\pm0.003$ ($5\times5$-grid). The minimal values of the displayed grids ($\sigma=0.0001$ for the $4\times4$-grid and $\sigma=0.002$ for the $5\times5$-grid) do not correspond to actual lower bounds, above which scaling behavior can be observed. The scaling regimes may extend down to $\sigma=0$, but it becomes numerically increasingly hard to identify the rather long periods. On the other hand, the upper values for $\sigma$, as displayed in Fig.~\ref{fig3}, do correspond to bounds in the sense that self-similarity of phase trajectories gets lost above these values. 

It should be furthermore noticed that the scaling exponents for both grids agree within the error bars, although the attractor space for both grids is supposed to be quite different: The $4\times4$-grid shows the WS-dimensional reduction with a continuum of attractors for $\sigma\rightarrow 0$, differently from the $5\times5$-grid. The reason for similar exponents may be a comparable value for the long period, which is 1524 t.u. for the $4\times4$-grid and 1485 for the $5\times5$-grid for the same choice of $\sigma$, here $\sigma=0.01$. When we considered  a long-period orbit with a period of $T=765$ on a $4\times4$-grid, again for $\sigma=0.01$, the scaling exponent was $b=1.336\pm 0.007$. However, for another pair of orbits on a $4\times4$ and $5\times5$-grid with comparable periods and the same $\sigma$ the exponents did not agree within the error bars, leading to the conclusion that the exponents of the scaling behavior are not universal.

\subsection{Rings with undirected couplings}
\begin{figure}[!t]
\centering{
\includegraphics[width=1.0\columnwidth]{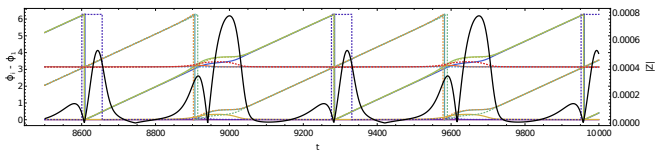}
}
\caption{An example of a long period orbit for a (6,12) ring and $\sigma=0.01$. The black curve shows the order parameter $\vert Z\vert$, while the colored curves represent the eleven phase differences $\phi_i-\phi_1$, where those belonging to the same cluster are hardly distinguishable in the figure. Here the system spends most of the time in a four-cluster state, and for a short time in a two or six-cluster state. }\label{fig4}
\end{figure}

Before we discuss self-similar temporary patterns on ring topologies, let us give an example for long-period orbits on a (6,12) ring with $\sigma=0.01$. The black curve in Fig.~\ref{fig4} shows the order parameter, while the colored ones represent the phase differences $\phi_i-\phi_1$. For this example the system spends most of the time in a four-cluster state, and for a short time in a two or six-cluster state. The four-cluster state consists of three oscillators each. Two of four clusters have a fixed phase difference $\pi$ between each other, let us call these clusters $c_{1}^1$ and $c_{2}^1$, while the other two $c_1^2$ and $c_2^2$ have a fixed phase difference between each other, but not to the clusters $c_i^1$. When the clusters $c_i^1$ and $c_j^2$ ``collide", the system spends a short time in a two or six-cluster state. Two-cluster states last for approximately 50 t.u., which corresponds to about 8 periods of an individual oscillator, and six-cluster states for about 150 t.u. or about 25 individual periods.

\begin{figure}[!t]
\centering{
\includegraphics[width=1.0\columnwidth]{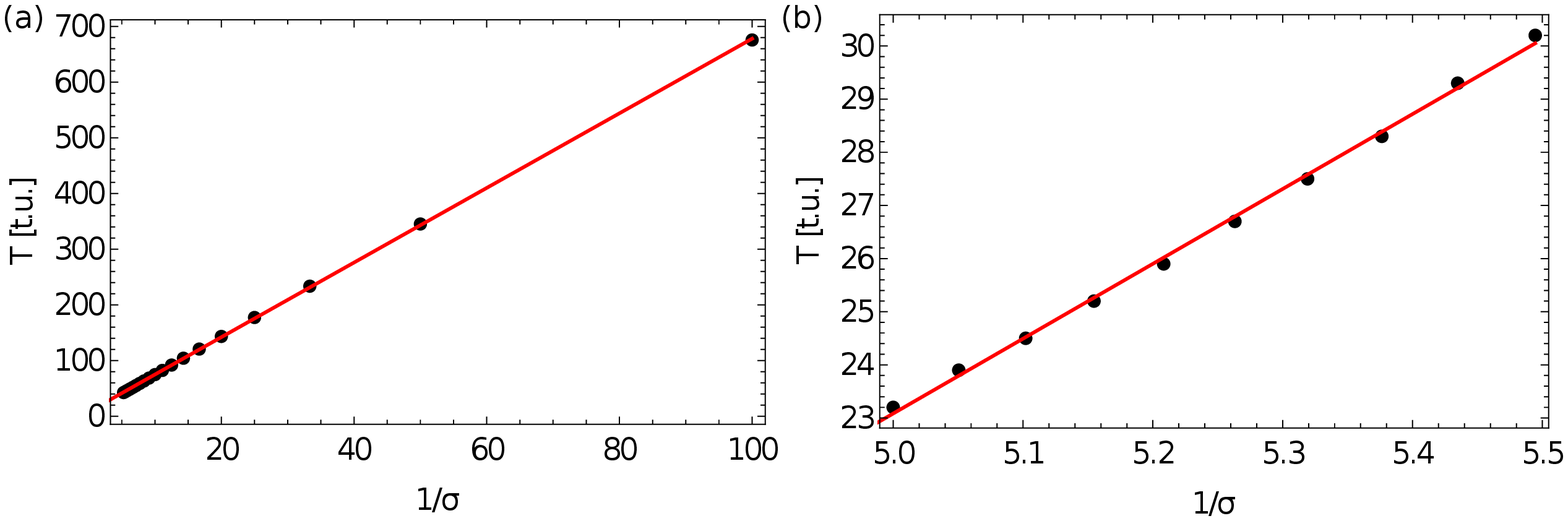}
}
\caption{Periods of the order parameter $\vert Z\vert$ along selfsimilar orbits as a function of the width $\sigma$ of the distribution of natural frequencies for a $(6,12)$-ring (a) and a $(4,15)$-ring (b).  The period $T$ scales linearly in $1/\sigma$. The fit (red line) is $T(\sigma) = a + b/\sigma$, where $a=8.1\pm 0.4$ and $b=6.70\pm0.01$ (a), and $a=-47\pm1$ and $b=14.1\pm0.2$ (b). The range of $\sigma$ is 0.01 to 0.19 (a) and 0.182 to 0.20 (b).}\label{fig5}
\end{figure}

Again, in Fig.~\ref{fig5} we display the scaling behavior of the order parameter period $\vert Z\vert$ of self-similar orbits as a function of $\sigma$, for a $(6,12)$-ring (a) and a $(4,15)$-ring (b). Both scaling behavior here fits best to $T(\sigma) = a + b/\sigma$, where $a=8.1\pm 0.4$ and $b=6.70\pm0.01$ (a), and $a=-47\pm1$ and $b=14.1\pm0.2$ (b). The range of the displayed $\sigma$-values is 0.01 to 0.19 (a) and 0.182 to 0.20 (b). For the $(4,15)$-ring, unlike for the other displayed cases, lower values for $\sigma$ are not omitted for accuracy reasons or exceeding numerical effort. The values of 0.182 corresponds to an actual lower bound, as we could not find any long-period orbits below $\sigma=0.182$ for 500 random initial conditions, so that the scaling regime here is bounded from above and below, it is more narrow than in the other considered cases. The linear scaling of $T$ in $1/\sigma$ seems to be in common.

\subsection{Stability of the long-period orbits}
One may wonder whether our long-period orbits play a role in natural systems, in particular in those, which can be reasonably approximated by the Kuramoto model. It should be noticed that the long lasting limit cycles here are not entrained  by external fields (differently from the circadian clock, where day-night or seasonal cycles entrain the suprachiasmatic nucleus). Here they emerge self-organized from the individual oscillations and consist of repeating patterns of phase-locked motion. Following the individual phase trajectories, the long periods may be easily overlooked, as on time scales shorter than a whole period the phase evolutions look quite irregular and chaotic.

In relation to neuronal networks and the conjectured and desired metastability  of the brain state,  a concept of chaotic itinerancy  was proposed in \cite{japan}. According to this concept, brain activity can be expressed  by a trajectory on an attractor which comprises unstable periodic orbits.
The process of perception then corresponds to a temporary stabilization of dynamics on such an orbit, before leaving it soon afterwards and continuing with free itinerancy.

\begin{figure}[!t]
\centering{
\includegraphics[width=.9\columnwidth]{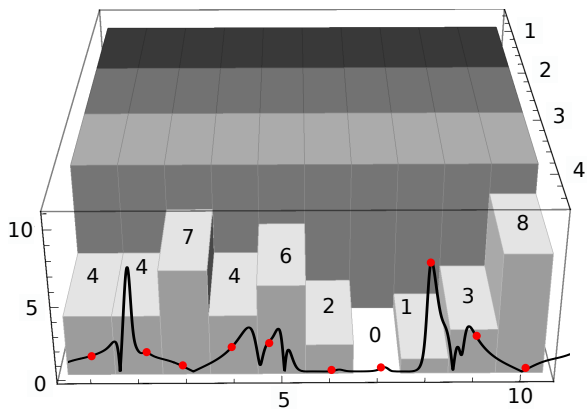}
}
\caption{Histogram of returns to the trajectory after a perturbation. The horizontal axis indicates time, the vertical axis the number of returns after 10 perturbations per choice of the radius, and the third axis represents 4 different distances $|\vec{v}|=\pi 10^{(i-4)}$, $i=1,..4$ from the main trajectory, chosen for the perturbed initial conditions (black representing the smallest perturbation, and lightest gray the largest).  The histogram shows that only the largest perturbations kick the system out of the basin of attraction of the initial periodic orbit.}\label{fig6}
\end{figure}

Therefore it is of interest to further quantify the stability of our long-period orbits. We perturb the trajectory in normal direction to the vector of motion in ten randomly chosen directions $\vec{v}$, and with four different values $|\vec{v}|=\pi 10^{(i-4)}$, $i=1,..4$ per direction. So the vectors of perturbations lie in 15-dimensional ``discs"  of different radii around the trajectory.
We let the system then evolve for $10^6$ t.u. and check whether it returns to the characteristics of the initial state in terms of the synchronization pattern, that is to the same value of $\vert Z\vert$ $10^6$ t.u. later. This procedure is applied for ten different instants of time (red dots in Fig.~\ref{fig6}), equidistantly distributed over approximately one period, see the black curve in  Fig.~\ref{fig6}, which represents $\vert Z\vert$. For the three smallest radii all 10 perturbed trajectories return to the trajectory of $\vert Z\vert$, but for the largest radius, between none and eight return, as indicated on top of the columns of the histogram.  So the size of the basin of attraction around the trajectory depends on the position in phase space, as one would also expect.

\section{Conclusions and outlook}
Although we cannot predict the observed scaling from a detailed bifurcation analysis, the very observations are rather interesting on their own. Often self-similarity  refers to static spatial patterns when zooming further into the spatial resolution and discovering fractal behavior. On the other hand, scale-invariance of  Hamiltonians like that of the Ising model at a second-order phase transition can be understood in terms of the renormalization group, when the relevant degrees of freedom are tuned towards the critical manifold and the couplings are renormalized, while the spatial scale is changed (see, for example, \cite{yeomans}). In the Ising  case, at the critical point, scale invariance of the Hamiltonian dynamics is manifest in clusters of aligned spins on all length scales. So the scale invariance of the Hamiltonian is manifest in self-similar static patterns of aligned spins. In contrast, we observed self-similarity of temporal sequences, when zooming in in time and readjusting the strength of the disorder. Again, this may be due to a dimensional reduction of the system to fewer effective degrees of freedom. In cases, for which we observed the dynamically generated WS-phenomenon, we know that at $\sigma=0$ the dimension of the dynamics drastically reduces to three degrees of freedom. However, as the scaling results for the $5\times 5$-grid without the WS-phenomenon indicate, the dimensional reduction need not be realized according to the WS-reduction. So it remains to identify whether in this case it is another effective reduction of the system's dimension that is responsible for the observed scaling.  As we have seen, the scaling of T with $\sigma^b$ is not universal for the two-dimensional grids. It certainly depends on the dimension of the system, $b=-1.0$ for the rings and $b < -1.0$ for the two-dimensional grids. Generic conditions for the scaling to happen and explaining it in terms of the  underlying attractor space remain a challenge for future research.

\section*{Acknowledgement}
We would  like to thank the organizers of the PNLD-2016 conference for a stimulating meeting from July 24-29 in Berlin. Financial support from Deutsche For- schungsgemeinschaft (DFG, contract ME-1332/25-1) is gratefully acknowledged.  We are also indebted to Michael Zaks (Humboldt University Berlin) for valuable discussions.



\end{document}